\documentclass[a4paper,12pt]{article}
\usepackage{amsmath}
\usepackage{graphicx}
\usepackage{rotating}
\usepackage{wasysym}
\DeclareGraphicsExtensions{.pdf,.png,.jpg}
 \def\dir{General_sty/}
\usepackage{chicago}
\usepackage{klett}
\usepackage{def_the}
\usepackage{Fig_size_A}
\usepackage{kfig}

\newtheorem{corollary}[theorem]{Corollary}
\newtheorem{assumption}[theorem]{Assumption}

\title{}
\author{Eckhard Platen}
\begin{document}
\thispagestyle{empty} \vspace*{1.0cm}

\begin{center}
{\LARGE\bf  Pricing  under the Benchmark Approach }
\end{center}

\vspace*{.5cm}
\begin{center}

{\large \renewcommand{\thefootnote}{\arabic{footnote}} {\bf Eckhard
Platen}\footnote{University of Technology Sydney,
  School of Mathematical and Physical Sciences, and \\Finance Discipline Group}$^{,}$
\vspace*{2.5cm}

\today
}

\end{center}

\begin{minipage}[t]{13cm}
The paper summarizes key results of the benchmark approach with a focus on the concept of benchmark-neutral pricing. It applies these results to the pricing of an extreme-maturity European put option on a well-diversified stock index.  The growth optimal portfolio  of the stocks  is approximated by a well-diversified stock portfolio and modeled by a drifted time-transformed squared Bessel process of dimension four. It is shown that the benchmark-neutral price of a European put option is theoretically the minimal possible price and the respective risk-neutral put price turns out to be significantly more expensive.

\end{minipage}
\vspace*{0.5cm}


\vspace*{0.5cm}
\vspace*{0.5cm}

\vspace*{0.5cm}

	\newpage
	\section{Introduction}\label{section.intro}
	After some work on pricing of  options under stochastic volatility, see, e.g., \citeN{HofmannPlSc92}, by applying the classical risk-neutral pricing methodology, 
	the author concluded that risk-neutral extreme-maturity  put option prices on a stock index,  with decades to maturity, appeared to be   overpriced. This insight  motivated the  development of  the benchmark approach; see \citeN{Platen02g}, \citeN{Platen06ba}, and \citeN{PlatenHe06}.\\  By removing assumptions of the classical mathematical finance theory, presented, e.g., in \citeN{Jarrow22}, the benchmark approach  gives access to a wider modeling world. With its concept of {\em real-world pricing}, the benchmark approach employs  the {\em growth optimal portfolio} (GOP) as a num\'eraire and the real-world probability measure as the respective pricing measure.  The GOP is   interchangeably  called the Kelly portfolio, expected logarithmic utility-maximizing portfolio, or num\'eraire portfolio; 
	see  \citeN{Kelly56}, \citeN{Merton71}, \citeN{Long90}, \citeN{Becherer01}, \citeN{Platen06ba}, \citeN{KaratzasKa07}, \citeN{HulleySc10}, and \citeN{MacLeanThZi11}. It maximizes the growth rate and is the  best-performing portfolio in the long run.\\
The main assumptions  for obtaining the real-world price of a contingent claim are extremely weak and consist of the existence of the GOP  and the existence of the real-world conditional expectation of the  contingent claim when denominated in the GOP; see \citeN{PlatenHe06} and \citeN{DuPl16}. Real-world pricing provides theoretically the minimal possible prices for a replicable contingent claim. \\
In practice, it is not easy to implement real-world pricing because the GOP 
of the entire financial market is a highly leveraged portfolio and is  difficult to approximate by a guaranteed strictly positive portfolio. Due to this difficulty, in \citeN{Platen25a} the concept of benchmark-neutral (BN) pricing was proposed. It employs the GOP of the market formed by the stocks (without the savings account) as a num\'eraire.  If the respective BN pricing measure is an equivalent probability measure, then the BN pricing formula provides the same minimal possible prices as the real-word pricing formula. In \citeN{Platen25a} it has been illustrated that for a realistic stock index model  extreme maturity bonds are significantly less expensive than those obtained by   risk-neutral pricing. Extreme maturity bonds are the typical building blocks of pension and life insurance products. Therefore, it matters for many people whether these prices are   more expensive than necessary. \\

	  The assumed  existence of the GOP     represents an extremely weak and easily verifiable assumption that the benchmark approach makes. It represents  a {\em no-arbitrage condition}   because \citeN{KaratzasKa07} and \citeN{KaratzasKa21} have shown that  the existence of the GOP is equivalent to their {\em No Unbounded Profit with Bounded Risk} (NUPBR) no-arbitrage condition. This no-arbitrage condition is weaker than the {\em No Free Lunch with Vanishing Risk} (NFLVR)  no-arbitrage condition of \citeN{DelbaenSc98}, which is underpinning the classical mathematical finance theory; see, e.g.,  \citeN{Jarrow22}.\\
	  
	  The current paper  illustrates BN pricing for an extreme maturity put option on a well-diversified stock portfolio.   It summarizes at first  theoretical results that underpin the benchmark approach and BN pricing  provided in \citeN{PlatenHe06}, \citeN{FilipovicPl09}, \citeN{DuPl16}, and \citeN{Platen25a}.  Later it employs the {\em minimal market model} (MMM), see \citeN{Platen01a}, by assuming that 
	 a well-diversified stock portfolio  follows  a time-transformed squared Bessel process of dimension four in some flexible activity time.  The resulting BN price of an extreme-maturity European put option on a well-diversified stock portfolio is  derived and illustrated using  historical data.  This price turns out to be significantly less expensive than the respective formally calculated risk-neutral price.

	    The paper is organized as follows:  Section 2  introduces  the  market setting and summarizes statements of the benchmark approach that will be applied  later.
	     Section 3  presents the concept of BN pricing.
	   Finally,   Section 4 prices a European put option under the MMM.
	       
	       \section{General Market Setting} \setcA \setcB 
	       This section summarizes  general results of the benchmark approach.
	       \subsection{Primary Security Accounts}

	        Consider a continuous market, which consists of a set of  $m\in \{1,2,,...\}$  {\em primary security accounts} $S^1_t,...,S^m_t$. 
	          When holding  units of a primary security account, its owner is reinvesting in this account all  interest payments, dividends,  or other payments  obtained or paid.
	       For matrices and vectors $ \bf{x}$ we 
	       denote by $ {\bf{x}}^\top$ their transpose.  Moreover, $\b1=(1, \dots, 1)^\top$ is a vector, and we write $\bf 0$ for a zero matrix or vector, where the dimensions follow from the context. 
	        Throughout the paper, the  modeling is performed on a filtered probability space $(\Omega,\mathcal{F},\underline{\cal{F}},P)$, satisfying the usual conditions; see, e.g., \citeN{Shiryaev84}, 
	       and \citeN{KaratzasSh98}.  The filtration $\underline{\cal{F}}$ $=(\mathcal{F}_t)_{t \in [0,\infty)}$ models the evolution of information relevant to the  market model. 
	       The information available at time $t \in [0,\infty)$ is captured by the sigma-algebra $\mathcal{F}_t$. \\ 
	       
	       The   vector   ${\bf{S}}_t=(S^1_t,..., S^{m}_t)^\top$ of the strictly positive  values of the primary security accounts at time  $t \in [0,\infty)$  is assumed to  be a unique strong solution of the $m$-dimensional  
	       It\^{o}-vector stochastic differential equation (SDE), see \citeN{ShiryaevCh02}, 
	       \begin{equation} \label{e.2.1}
	       \frac{d{\bf{S}}_t}{{\bf{S}}_t}=\mu_t dt +
	       {\bf{\sigma}}^{}_t d{\bf{W}}_t
	       \end{equation}
	       	       with vector of strictly positive {\em initial values}  ${\bf{S}}_0$ and  a vector process ${\bf{W}}=\{ {\bf{W}}_t =(W^1_t,\dots,W^{n}_t)^\top,t \in [0,\infty)\}$ of  $n\in\{1,2,...\}$  driving independent  standard {\em Brownian motions}. 
	        The dynamics of the primary security accounts are determined by the  {\em expected return} vector $\mu_t=(\mu^{1}_t,\dots,\mu^{m}_t)^\top$, and the  {\em volatility} matrix   ${\bf{\sigma}}_t=[\sigma^{j,k}_t]_{j,k=1}^{m,n}$ for   $t \in [0,\infty)$. Here we write $\frac{d{\bf{S}}_t}{{\bf{S}}_t}$ for the $n$-vector of It\^o-stochastic differentials $(\frac{dS^1_t}{S^1_t},...,\frac{dS^{m}_t}{S^{m}_t})^\top$.

	       The components of the  processes ${\bf \mu}
	       $ and 
	       $\sigma$, as well as other quantities and processes that  will be employed, 
	       are assumed to have 
	       appropriate measurability and integrability properties so that all manipulations we will perform are permitted. In particular, the volatility matrix components are assumed to be predictable and square integrable. The components of the expected return vector are assumed to be predictable and integrable.\\
	      
	       \subsection{Portfolios}
	       
	      The market participants can form portfolios by combining primary security accounts.  If changes in the value of a portfolio are only due to changes in the values of the
	      primary security accounts, then no extra funds flow in or out of the portfolio, and
	      the corresponding portfolio and strategy are called {\em self-financing}. A positive  self-financing {\em portfolio} $S^\pi_t$  is described by its  initial value $S^\pi_0>0$ and its  {\em weight process} $\pi=\{\pi_t=( \pi^1_t,..., \pi^{m}_t)^\top,t\in[0,\infty)\}$. The latter is describing the 
	       fractions of wealth invested in the respective securities, where
	       \begin{equation}\label{sumpi}
	       \pi_t^\top {\bf 1}=1
	       \end{equation}  for $t \in [0,\infty)$. 
	        The value of a self-financing portfolio  $S^\pi_t$ satisfies the SDE
	       \begin{equation} \label{e.2.12}
	       \frac{dS^\pi_t}{S^\pi_t}=\pi_t^\top\frac{d{\bf{S}}_t}{{\bf{S}}_t}=\pi_t^\top \mu_t dt +\pi_t^\top 
	       {\bf{\sigma}}^{}_t d{\bf{W}}_t
	       \end{equation}
	      	       for $t \in [0,\infty)$. By application of the It\^o formula, one can show that a portfolio that is self-financing in the denomination of one price process is also self-financing in the denomination of other price processes. If not otherwise stated, we consider self-financing portfolios.\\
	       
	        We introduce for a strictly positive portfolio $S^\pi_t$ its {\em strategy}  $\delta = \{\delta_t = (\delta^1_t,...,\\ \delta^m_t)^\top, t\in[0,\infty)\}$, where 
	        \begin{equation}\delta^j_t=\frac{\pi^j_t S^\pi_t}{S^j_t}\end{equation}  represents the number of units of the $j$-th primary security account that are held at time $t\in[0,\infty)$ in the
	         portfolio for $j\in\{1,...,m\}$. 
	        The self-financing property of a  portfolio is expressed by the equation
	        \begin{equation}\label{stochint}
	         S^\pi_t= S^\pi_{0}+\int_{0}^{t}\delta_s^\top d{\bf{S}}_s
	        \end{equation}
	        with \begin{equation}  S^\pi_{t}=\delta_{t}^\top{\bf{\ S}}_{t}
	        \end{equation}
	         for $t\in[0,\infty)$, where the stochastic integral in \eqref{stochint} is assumed to be a vector-It\^{o} integral; see \citeN{ShiryaevCh02}. \\
	        To
	        introduce a class of admissible strategies for forming
	        portfolios, denote by $[{\bf{ S_.}}]_t=([ S^i_., S^j_.]_t)_{i,j=1}^{m} $ the matrix-valued optional covariation  of the vector  of  primary security
	        accounts ${\bf{ S}}_t$ for  $t\in[0,\infty)$.

	        \begin{definition}\label{integr'}
	        	An {\em admissible self-financing strategy} ${\bf {\delta}}=\{\delta_t=(\delta^1_t,...,\delta^m_t)^\top,t\in[0,\infty)\}$, initiated at the time $ 0$, is an ${\bf{R^{m}}}$-valued, predictable
	        	stochastic process, satisfying the condition
	        	\begin{equation}\label{integrability'}
	        	\int_{0}^{t}\delta^\top_u[{\bf{ S_.}}]_u \delta_u du<\infty
	        	\end{equation} 
	        	for  $t\in[0,\infty)$.
	        \end{definition}
	       
	     \subsection{Growth Optimal Portfolio}   
	        The following portfolio  plays a central role  in the benchmark approach:
	       \begin{definition}
	       For the given  market, a {\em growth optimal portfolio} (GOP)  is defined as a strictly positive self-financing portfolio $S^\pi_t$ with initial value $S^\pi_0=1$ that maximizes at all times $t \in [0,\infty)$  its instantaneous growth rate 
	       \begin{equation}\label{gpi}
	       g^\pi_t=\pi^\top_t  \mu_t-\frac{1}{2}\pi^\top_t  \sigma_t  \sigma^\top_t  \pi_t.
	       \end{equation}
	       \end{definition} 
	       This leads  to the  $m$-dimensional constrained quadratic optimization problem
	       \begin{equation}\label{quop}
	       \max \left\{g^\pi_t| \pi_t \in {\bf R}^{m}, \pi^\top_t  {\bf 1}=1\right\}
	       \end{equation}
	       for $t \in [0,\infty)$.	 To characterize its solution, we introduce the symmetric 
	       matrix \BE  {\bf M}_t=\left( \begin{array}{cc} \sigma_t \sigma^\top_t & \b1  \\ \b1^\top & 0 \end{array} \right) 
	       \EE
	       and denote by $im\left(	{\bf M}_t\right) $ the image, range, or column space of this matrix; see, e.g., \citeN{GolubVa96}. 	       
	        We say that a GOP {\em exists} for the given market if the above optimization problem has a  solution.    
	        Theorem 3.1 in \citeN{FilipovicPl09} provides  the following general statements:       
	       \begin{theorem}
	       	\label{GOPTheorem}
	       A GOP $S^{*}_t$ exists for the above continuous market if and only if 
	       \begin{equation}\label{imM}
	       \left( \begin{array}{c} \mu_t \\ 1 \end{array} \right)\in im\left(	{\bf M}_t\right) 
	       \end{equation}
	       for all $t \in [0,\infty)$. In the case when the above condition is satisfied, the following statements emerge:  The value process of the GOP $S^{*}_t$ is unique and satisfies the SDE
	       \BE \label{e.4.1}
	       \frac{dS^{*}_t}{S^{*}_t}=\lambda^{*}_t dt +
	       ({ \theta_t})^\top  ( { \theta_t} dt+d{\bf W}_t) \EE
	       for $t \in [0,\infty)$ with $S^{*}_0=1$, where its optimal weight vector  $\pi^*_t=(\pi^{*,1}_t,...,\pi^{*,m}_t)^\top$ together with its  {\em generalized risk-adjusted return} $\lambda^{*}_t$ 
	       represent  a solution to the  equation
	       \BE \label{matrixequation} \left( \begin{array}{cc} \sigma_t \sigma^\top_t & \b1  \\ \b1^\top & 0 \end{array} \right) \left( \begin{array}{c} \pi^*_t \\ \lambda^{*}_t 
	       \end{array} \right)=\left( \begin{array}{c} \mu_t \\ 1 \end{array} \right), \EE
	       	       with the vector of {\em generalized market prices of risk}
	       	       \BE \label{e.4.2} { \theta_t}=(\theta^{1}_t,...,\theta^{n}_t)^\top=\sigma^\top_t  \pi^*_t
	       \EE
	      	      and 
	      \begin{equation}\label{sumpi1}
	     ( \pi^*_t)^\top{\bf {1}}=1
	      \end{equation} for $t \in [0,\infty)$. The generalized risk-adjusted return $\lambda^*_t$ is zero when the denominating process is a  portfolio of traded primary security accounts. The value of a strictly positive self-financing portfolio $S^\pi_t$ with weight process $\pi$ satisfies the SDE
	       \begin{equation}\label{portfolio}
	       \frac{dS^{\pi}_t}{S^{\pi}_t}=\lambda^{*}_t dt +\pi_t^\top
	       \sigma_t  (  \theta_t dt+d{\bf W}_t) \end{equation}
	      with   generalized  risk-adjusted return $\lambda^{*}_t$ satisfying the vector equation
	      \begin{equation}\label{lambda*'}
	      \lambda^{*}_t=\pi_t^\top\mu_t-\pi_t^\top\sigma_t  \theta_t, 
	      \end{equation} and its  value $\frac{S^\pi_t}{S^{*}_t}$ denominated in the GOP satisfies the driftless  SDE
	       \begin{equation}\label{bmportfolio}
	       \frac{d\frac{S^\pi_t}{S^{*}_t}}{\frac{S^\pi_t}{S^{*}_t}}={\bf u}^\pi_td{\bf W}_t \end{equation}
	       with volatility
	       \begin{equation}\label{bmportfolio'}
	       {\bf u}^\pi_t=\pi_t^\top
	       \sigma_t- \theta_t^\top  \end{equation}
	      	       for $t \in [0,\infty)$. 
	      \end{theorem} 
	      The above theorem provides necessary and sufficient conditions for the existence of a GOP. It reveals the natural structure of any security or portfolio. These have  the same generalized risk-adjusted return $\lambda^*_t$ and the same generalized market prices of risk $\theta_t$. All securities and portfolios, when denominated in the GOP satisfy driftless SDEs, which means that they are local martingales. 
	       
	      \subsection{Long-Term Growth Rate}
	      The GOP is a  special portfolio, in respect to its long-run performance. Let us define for a strictly positive portfolio process $S^\pi=\{S^\pi_t,t\in[0,\infty)\}$ its {\em long-run growth rate} as the almost sure limit
	      \begin{equation}\label{gpi1}
	      g^\pi=\limsup_{T\rightarrow \infty}\frac{1}{T}\ln\left(\frac{S^\pi_T}{S^\pi_0}\right)
	      \end{equation}
	    and  assume that this limit exists for the GOP.   The fascinating  property of the GOP is that  its trajectory attains, in the long run, almost surely pathwise  values not less than those of any other strictly positive portfolio. This property is captured in the following result, which  is derived in the proof of Theorem 10.5.1 in \citeN{PlatenHe06}: 
	      \begin{theorem}
	      	\label{longtermgr}
	      	For the in Theorem \ref{GOPTheorem} given market, it holds $P$-almost surely for the long-run growth rate of all strictly positive portfolios $S^\pi_t$ the inequality\begin{equation}
	      	g^{\pi^*}\geq g^\pi,
	      	\end{equation}
	      	where $\pi^*$ describes the GOP strategy.
	      \end{theorem}

	       \subsection{Locally Risk-Free Portfolio}

	       A {\em locally risk-free portfolio} (LRP) $S^0_t$ is a portfolio with zero volatility and initial value $S^0_0=1$ that is given in the form
	       \begin{equation}\label {A0}
	       S^0_t=\exp\left\{\int_{0}^{t}r_sds\right\}
	       \end{equation}
	       for  $t \in [0,\infty)$, where $r=\{r_t,t \in [0,\infty)\}$ denotes the LRP's adapted {\em risk-free return} process.
	       Let us denote by $	\ker(\sigma_t)$ the kernel or nullspace of a matrix or vector $\sigma$; see, e.g., \citeN{GolubVa96}. According to Lemma 4.1 and Theorem 4.1 in \citeN{FilipovicPl09}   the following general statements hold:
	       \begin{theorem}\label{MVPTheorem}
	       	For the  market given in Theorem \ref{GOPTheorem}, an LRP $S^0_t$ exists if and only if 
	       	\begin{equation}\label{ker2}
	       	\ker((\sigma_t)^\top)\notin \ker({\bf{1}}^\top) 
	       	\end{equation}
	       	for all $t \in [0,\infty)$. If an LRP exists, then the risk-free return equals
	       	\begin{equation}
	       	r_t=(\pi^{LRP}_t)^\top\mu_t
	       	\end{equation}
	       	for some weight vector $\pi^{LRP}_t \in \ker((\sigma_t)^\top)$ with  $(\pi^{LRP}_t)^\top {\bf {1}}=1$  for $t \in [0,\infty)$. If an LRP and the GOP exist in the given market, then one has
	       	\begin{equation}\label{rtlambda}
	       	r_t=\lambda^{*}_t
	       	\end{equation}
	       	for $t \in [0,\infty)$.
	       \end{theorem}

	       \subsection{Market Extensions}
	       	       When modeling a  market, the question may arise whether one can add  a particular security to this market. This leads to the question concerning the GOP of the extended market. The following three-fund separation theorem, which is derived as Theorem 7.1 in  \citeN{FilipovicPl09},  answers this   question: 
	       \begin{theorem}
	       	\label{extension}
	       	When extending the in Theorem \ref{GOPTheorem} given market of $m$ primary security accounts 
	       	 by a new  security account $ \Sigma_t$, satisfying the SDE
	       	\begin{equation}
	       	\frac{d \Sigma_t}{ \Sigma_t}=\alpha_tdt+\beta_t^\top d{\bf {W}}_t
	       	\end{equation} 
	       	for $t\in[0,\infty)$ with $\Sigma_0>0$,
	       	the GOP $ S^{**}_t$
	       	for the extended market with primary security accounts $S^1_t,...,S^m_t, \Sigma_t$ exists
	       	if and only if for all $t\in[0,\infty)$ at least one of the following two conditions holds:
	       	\begin{equation}\label{alphat}
	    (i)\hspace{3.2cm}   	\alpha_t=\lambda^*_t+\beta^\top_t \theta_t
	       	\end{equation}
	       	or
	       	\begin{equation}\label{imM2}
	       (ii)\hspace{2.0cm} 	\left( \begin{array}{c} \sigma_t\beta_t \\ 1 \end{array} \right)\in im\left(	{\bf M}_t\right). 
	       	\end{equation}
	       	In  case $(ii)$,  a strategy for the GOP $ S^{**}_t$ of the extended market is given by the three-fund separation
	       	\begin{equation}
	       	\pi^{**}_t= \left( \begin{array}{c} \pi^*_t \\ 0 \end{array} \right)+ p^*_t\left( \begin{array}{c} {\bf 0} \\ 1 \end{array} \right)-p^*_t \left( \begin{array}{c} {\bf x}^*_t \\ 0 \end{array} \right)
	       	\end{equation}
	       	with unique extended GOP value $ S^{**}_t$ satisfying the SDE
	       	\BE \label{e.4.20}
	       	\frac{d S^{**}_t}{ S^{**}_t}=\lambda^{**}_t dt +
	       	(  \theta^{**}_t)^\top  (   \theta^{**}_t dt+d{\bf W}_t), \EE	
	       	where
	       	\begin{equation}
	       	\lambda^{**}_t=\lambda^*_t-p^*_t(\beta_t-\sigma_t^\top {\bf x}^*_t)^\top\sigma_t^\top {\bf x}^*_t,
	       	\end{equation}
	       	\begin{equation}
	       	\theta^{**}_t=\theta_t+p^*_t(\beta_t-\sigma_t^\top {\bf x}^*_t)^\top,
	       	\end{equation}
	       	and $(({\bf x}^*_t)^\top,p^*_t)\in {\bf R}^m$ are uniquely determined by $\mu_t, \sigma_t, \alpha_t$,  and $\beta_t$.\\ If case $i$ holds, then 
	       	\begin{equation}
	       	(({\bf x}^*_t)^\top,p^*_t)={\bf 0},
	       	\end{equation}
	       	and if case (ii) holds, then  ${\bf x}^*_t$
	       	is a solution of the well-posed minimization problem
	       	\begin{equation}\label{quop1}
	       	\min \left\{|\beta_t-\sigma_t^\top {\bf x}_t|^2| {\bf x}_t \in {\bf R}^{m}, ({\bf x}_t)^\top  {\bf 1}=1\right\},
	       	\end{equation}
	       	with first order conditions
	       	\BE  \left( \begin{array}{cc} \sigma_t \sigma^\top_t & \b1  \\ \b1^\top & 0 \end{array} \right) \left( \begin{array}{c} {\bf x}_t \\ p_t 
	       	\end{array} \right)=\left( \begin{array}{c} \sigma_t\beta_t \\ 1 \end{array} \right), \EE
	       	and $p^*_t$ is determined by the equation
	       	\begin{equation}
	       	p^*_t   |\beta_t-\sigma_t^\top {\bf x}^*_t|^2=\alpha_t -\lambda^*_t+\beta_t^\top\theta_t.
	       	\end{equation}
	       	Therefore, $\beta_t=\sigma_t^\top {\bf x}^*_t$ necessitates the case $(i)$.       \end{theorem}
	       One has the following interpretation of the above theorem: Suppose that the case $(ii)$ holds, then there exists a portfolio $S^{x^*}_t$ in the original market that optimally replicates the new security account $\Sigma_t$ in the sense that
	       it minimizes the instantaneous conditional variance $|\beta_t-\sigma_t^\top {\bf x}^*_t|^2$
	       of its unhedgeable
	       return component $\frac{d \Sigma_t}{\Sigma_t}- \frac{d S^{x^*}_t}{S^{x^*}_t}$, see \eqref{quop1}. The  GOP $S^{**}_t$ of the extended market is then obtained by
	       investing in the  GOP $S^*_t$ of the original market and holding a long (short) position $p^*_t$ in the new security
	       account $\Sigma_t$, balanced by a short (long) position $-p^*_t$
	       in the portfolio $S^{x^*}_t$. If case $(i)$ holds, then adding $\Sigma_t$ to the original market does not cause a GOP that is different from the one of the original market.

	   	\section{Benchmark-Neutral Pricing}\setcA
	    This section  introduces the concept of benchmark-neutral (BN) pricing for a simplified market with $n=1$ driving Brownian motion as a source of randomness.
	   	\subsection{GOPs}
	   	  In the following, we consider a simple market by assuming $m=1$
	   	 nonnegative  {\em risky primary security account}, denoted by $S^1_t$, 
	   	where all  dividends  are reinvested. Furthermore, we call this  primary security account $S^1_t$ the stock GOP $S^*_t=S^1_t$. 
	   	If not otherwise stated, we assume the securities  to be denominated   in units of the LRP $S^0_t=1$. 
	   	We emphasize that  the LRP $S^0_t$ is not included in the market given by the single primary security account  $S^1_t=S^*_t$.\\ 
	   	The stock GOP $S^*_t$ in savings account denomination is assumed to be continuous and to satisfy according to equation \eqref{e.4.1} in
	   	Theorem \ref{GOPTheorem}  the SDE
	   	\begin{equation}\label{dS*T}
	   	\frac{dS^{*}_t}{S^{*}_t}=\lambda^*_t dt +
	   	\theta_t  (  \theta_t dt+d W_t) \end{equation}
	   	for $t\in[0,\infty)$ with $S^*_{0}>0$.
	   	Here, $\lambda^*_t$ denotes the {\em generalized risk-adjusted return} of  $S^*_t$,  $\theta_t$ its {\em volatility}, and 
	   	$W=\{W_t, t\in[0,\infty)\}$  a one-dimensional $(P, \mathcal{\underline F})$-Brownian motion. 
	   		   	\\
	   	When we   add the savings account $S^0_t=1$ as an additional primary security account to the market formed by the single primary security account $S^1_t=S^*_t$,  the GOP     $S^{**}_t$  of the resulting market exists, and it follows directly from the case $(ii)$  
	   of	Theorem \ref{extension} 
	   	 that the GOP  $S^{**}_t$ of the extended market  satisfies the SDE 
	   	\begin{equation} \label{e.4.2'}
	   	\frac{dS^{**}_t}{S^{**}_t}=
	   	\sigma^{**}_t(  \sigma^{**}_t dt+d W_t) \end{equation}
	   	with initial value $S^{**}_{0}=1$ and {\em market price of risk}
	   	\begin{equation}\label{sigma**}
	   	\sigma^{**}_t=\frac{\lambda^*_t   }{\theta_t}+\theta_t
	   	\end{equation} 
	   	for $t\in[0,\infty)$. 
	   	\subsection{Real-World Pricing}
	   	We denote by ${\bf E}^{P}(.|\mathcal{F}_t)$  the conditional expectation under the real-world probability measure $P$, conditional on the information $\mathcal{F}_t$ available at time $t$.	Consider a bounded stopping time $T > 0$, and let $\mathcal{L}^1(\mathcal{F}_T)$
	   	denote the set of integrable, $\mathcal{F}_T$-measurable random variables in the given filtered probability space.
	   	\begin{definition}\label{def2.1}
	   		For a bounded stopping time $T \in(0,\infty)$, a 
	   		nonnegative payoff $H_T$, denominated in units of the savings account is called a {\em contingent claim} if $\frac{H_T}{S^{**}_T} \in
	   		\mathcal{L}^1(\mathcal{F}_T)$.
	   	\end{definition}
	   	As shown in Section 10.4 in \citeN{PlatenHe06} and  Corollary 6.2 in \citeN{DuPl16},
	   	for a
	   	contingent claim $\ H_T$ with maturity at a bounded stopping time $T$ the {\em real-world pricing formula}
	   	\begin{equation}\label{RWPF}
	   	H_t=S^{**}_t{\bf E}^{P}\left(\frac{{H}_{T}}{S^{**}_T}|\mathcal{F}_t\right)\end{equation} determines its, so-called, {\em fair}  price $ H_t$ for all $t\in[0,T]$. The ratio $\frac{H_t}{S^{**}_t}$ forms a $(P, \mathcal{\underline F})$-martingale, which is $P$-almost surely unique. 
	   		The real-world pricing formula uses the GOP $S^{**}_t$ of the extended market  as num\'eraire and the real-world probability measure $P$ as a pricing measure.  By Corollary 6.2 in \citeN{DuPl16}, for a replicable contingent claim,  its fair price process coincides with the value process of the minimal possible self-financing hedge portfolio that replicates the contingent claim. 
	  
	   	The pricing and hedging of nonreplicable contingent claims is studied in  \citeN{DuPl16} by employing the notion of benchmarked risk-minimization.

	   	\subsection{Change of Num\'eraire} 
	   	The 
	   	 GOP $S^{**}_t$ of the entire financial market is, in reality, a highly leveraged portfolio that goes    short in the savings account $S^0_t$; see Theorem \ref{extension}. When  hedging  contingent claims, one needs to be able to trade  the num\'eraire, for instance, when hedging a zero-coupon bond. 
	   A proxy of a  leveraged portfolio  can only be traded at discrete  times and, therefore,  faces  the possibility of becoming negative.  
	   	To avoid the above-mentioned difficulties, the paper \citeN{Platen25a} suggests  employing the 
	   	stock GOP $S^*_t$ as a num\'eraire and calls the respective pricing method  {\em benchmark-neutral  pricing} (BN pricing). 
	   	
	   	\noindent  As in  \citeN{Platen25a} the current paper makes the following assumption, which will be satisfied  for  the  stock GOP  model that we will employ later for  the illustration of BN pricing:
	   	\begin{assumption}\label{RNDMARTINGALE}
	   		The stock GOP $S^*_t$, when denominated  in units of the GOP $S^{**}_t$ of the extended market, forms the  $(P, \mathcal{\underline F})$-martingale $\frac{S^*}{S^{**}}=\{\frac{ S^*_t}{S^{**}_t},t\in[0,\infty)\}$, where
	   		$\frac{ S^*_t}{S^{**}_t}$	is the unique strong solution of the SDE
	   		\begin{equation}
	   		\frac{d(\frac{ S^*_t}{S^{**}_t})}{\frac{ S^*_t}{S^{**}_t}}=-\sigma^{S^*}(t)dW_t
	   		\end{equation}
	   		with finite integrated squared volatility
	   		\begin{equation*}
	   		\int_{0}^{t}\sigma^{S^*}(s)^2ds<\infty
	   		\end{equation*} for $t\in[0,\infty)$. 
	   	\end{assumption}

	    This leads us to  the  Radon-Nikodym derivative \begin{equation}\label{RND}
	   	\Lambda_{S^*}(t)=\frac{dQ_{S^*}}{dP}\Bigg\vert_{\mathcal{F}_t}=\frac{\frac{ S^*_t}{S^{**}_t}}{\frac{ S^*_{0}}{S^{**}_{0}}}=\exp\left\{-\int_{0}^{t}\sigma^{S^*}(s)dW_s-\frac{1}{2}\int_{0}^{t}\sigma^{S^*}(s)^2ds\right\},\end{equation} which characterizes for the num\'eraire $S^*_t$ the respective {\em BN pricing measure} $Q_{S^*}$; see \citeN{GemanElRo95}.
	   	By application of the It\^{o} formula we obtain from \eqref{dS*T},  \eqref{e.4.2'}, and \eqref{sigma**} 
	   		under the Assumption \ref{RNDMARTINGALE} for $\Lambda_{S^*}$  the SDE
	   		\begin{equation}
	   		\frac{d\Lambda_{S^*}(t)}{\Lambda_{S^*}(t)}
	   		=-\sigma^{S^*}(t)dW_t
	   		\end{equation}
	   		with
	   		\begin{equation}\label{sigma10}
	   		\sigma^{S^*}(t)=\sigma^{**}_t-\theta_t=\frac{\lambda^*_t}{\theta_t}
	   		\end{equation}
	   		for $t\in[0,\infty)$.  Furthermore,	   	
	   	 for  $T\in[0,\infty)$ we have for an event $A\in \mathcal{ F}_T$ its BN probability \begin{equation}Q_{S^*}(A)={\bf{E}}^{Q_{S^*}}({\bf{1}}_A)={\bf{E}}^P(\Lambda_{S^*}(t){\bf{1}}_A).\end{equation}
	   	Here ${\bf{1}}_A$ denotes the indicator function of the event $A$ and ${\bf{E}}^{Q_{S^*}}(.)$ the expectation under $Q_{S^*}$. Two measures are called {\em equivalent} if they have the same sets of events of measure zero. We say that a measure is an {\em  equivalent probability measure} if it is equivalent to the real-world probability measure $P$.\\	As a self-financing portfolio that is denominated in units of the GOP $S^{**}_t$ of the extended market, the Radon-Nikodym derivative $\Lambda_{S^*}(t)$ forms  a $(P, \mathcal{\underline F})$-local martingale. Under Assumption \ref{RNDMARTINGALE}  it is assumed to be a true martingale.  Let  ${\bf E}^{Q_{S^*}}(.|\mathcal{F}_t)$ denote  the conditional expectation  under the BN pricing measure $Q_{S^*}$ for the information available at time $t\in[0,\infty)$.  By Theorem 9.5.1 in \citeN{PlatenHe06} one obtains  the following Bayes' rule:
	   	\begin{corollary}\label{Theorem3.2}
	   		Under Assumption \ref{RNDMARTINGALE}, for some bounded  stopping time $T\in[0,\infty)$ and any $\mathcal{ F}_T$-measurable nonnegative random variable $Y=\frac{S^*_{0}H_T}{S^*_T}$, satisfying the integrability condition ${\bf{E}}^{Q_{S^*}}(Y)<\infty$, one has the Bayes' rule
	   		\begin{equation}
	   		{\bf{E}}^{Q_{S^*}}(Y|\mathcal{ F}_s)=\frac{{\bf{E}}^P(\Lambda_{S^*}(T)Y|\mathcal{ F}_s)}{{\bf{E}}^P(\Lambda_{S^*}(T)|\mathcal{ F}_s)}
	   		\end{equation}
	   		for $s\in[0,T]$ and, therefore, 
	   		\begin{equation}
	   		{\bf{E}}^{Q_{S^*}}\left(\frac{S^*_{0}H_T}{S^*_T}|\mathcal{ F}_{0}\right)={\bf{E}}^P\left(\frac{S^{**}_{0}H_T}{S^{**}_T}|\mathcal{ F}_{0}\right).
	   		\end{equation}
	   	\end{corollary} 
	   	  From  Corollary \ref{Theorem3.2} and Theorem 9.5.2 in \citeN{PlatenHe06} the following BN pricing formula and a BN version of  Girsanov's Theorem are obtained in  \citeN{Platen25a}:

	   	\begin{theorem}
	   		Under the Assumption \ref{RNDMARTINGALE}, for some bounded stopping time $T\in[0,\infty)$ and an $\mathcal{ F}_T$-measurable contingent claim $H_T$, satisfying the integrability condition ${\bf{E}}^{Q_{S^*}}(\frac{H_{T}}{S^*_T})<\infty$, the BN pricing measure $Q_{S^*}$ is an equivalent probability measure. The fair price $H_t$ at time $t\in[0,T]$, which the real-world pricing formula identifies for  $ H_T$, is  obtained via the {\em BN pricing formula} 
	   		\BE \label{BPF}\frac{H_t}{S^*_t} = {\bf E}^{Q_{S^*}}\left(\frac{H_{T}}{S^*_T}|\mathcal{F}_t\right) \EE
	   		for $t\in[0,T]$.   The process $\bar W=\{\bar W_t, t\in[0,\infty)\}$,
	   		satisfying the SDE
	   		\begin{equation}\label{W0}
	   		d\bar W_t=\sigma^{S^*}(t)dt+dW_t
	   		\end{equation}
	   		for $t\in[0,\infty)$ with $\bar W_{0}=0$, is under  $Q_{S^*}$ a Brownian motion. 
	   	\end{theorem}
	   We will see later  that BN pricing is capturing the stock GOP dynamics under $Q_{S^*}$  as if under $P$ the net risk-adjusted return $\lambda^*_t$ were zero. 

	   	\subsection{Admissible Strategies}
	   	The market participants can combine primary security accounts to form portfolios.
	   	Denote by $\delta = \{\delta_t = (\delta^0_t,\delta^1_t)^\top, t\in[0,\infty)\}$
	   	the respective portfolio strategy, where $\delta^j_t$, $j\in\{0,1\}$, represents the number of units of the $j$-th primary security account that are held at time $t\in[0,\infty)$ in a
	   	corresponding portfolio $S^\delta_t$. 
	   	To
	   	introduce a class of admissible strategies for forming
	   	portfolios, denote by $[{\bf{\tilde S_.}}]_t=([\tilde S^i_.,\tilde S^j_.]_t)_{i,j=0}^{1} $ the matrix-valued optional covariation  of the vector  of stock GOP-denominated primary security
	   	accounts ${\bf{\tilde S}}_t$ for  $t\in[0,\infty)$.

	   	\begin{definition}\label{integr'}
	   		An {\em admissible self-financing strategy} ${\bf {\delta}}=\{\delta_t=(\delta^0_t,\delta^1_t)^\top,\\t\in[0,\infty)\}$, initiated at the time $ 0$, is an ${\bf{R^{2}}}$-valued, predictable
	   		stochastic process, satisfying the condition
	   		\begin{equation}\label{integrability'}
	   		\int_{0}^{t}\delta^\top_u[{\bf{\tilde S_.}}]_u \delta_u du<\infty
	   		\end{equation} 
	   		for  $t\in[0,\infty)$.
	   	\end{definition}

	   	\subsection{ BN-Replicable Contingent Claims}
	   	In the following, we consider contingent claims that can be  replicated by using self-financing portfolios under BN pricing.  For a bounded stopping time $T$ let the set $ \mathcal{L}^1_{Q_{S^*}}(\mathcal{F}_T)$ denote the set of $\mathcal{F}_T$-measurable and $Q_{S^*}$-integrable random variables in the filtered probability space  $(\Omega,\mathcal{F},\underline{\cal{F}},Q_{S^*})$.\\
	   	
	   	\begin{definition}\label{HT}
	   		We call for a bounded stopping time $T\in[0,\infty)$ a stock GOP-denominated contingent claim  $\tilde H_T=\frac{H_T}{S^*_T}\in \mathcal{L}^1_{Q_{S^*}}(\mathcal{F}_T)$ {\em BN-replicable}   if it
	   		has for all $t\in[0,T]$ a representation of the  form
	   		\begin{equation}\label{tildeHT}
	   		\tilde H_T={\bf E}^{Q_{S^*}}(\tilde H_{T}|\mathcal{F}_t)+\int_{t}^{T}\delta_{\tilde H_T}(s)^\top d{\bf{\tilde S_s}}
	   		\end{equation}
	   		$Q_{S^*}$-almost surely, involving some predictable vector process $\delta_{\tilde H_T}=\{\delta_{\tilde H_T}(t)=(\delta_{\tilde H_T}^0(t),
	   		\delta_{\tilde H_T}^1(t))^\top,t\in[0,T]\}$  satisfying the condition \eqref{integrability'}.
	   	\end{definition}
	   	To capture the replication of a targeted contingent claim we introduce the following notion:
	   	\begin{definition} \label{deliver}
	   		An admissible self-financing  strategy $\delta=\{\delta_t=(\delta^0_t, \delta_t^1)^\top,t\in[0,T]\}$ is said to {\em deliver} the stock GOP-denominated BN-replicable contingent claim $\tilde H_T$ at a bounded stopping time $T$ if  the equality
	   		\begin{equation}
	   		\tilde S^\delta_T=\tilde H_T
	   		\end{equation}
	   		holds $Q_{S^*}$-almost surely.
	   	\end{definition}
	   	By combining  the above definitions, the following conclusion is stated in \citeN{Platen25a}:
	   	\begin{corollary}
	   		For a BN-replicable stock GOP-denominated contingent claim $\tilde H_T$ with representation \eqref{tildeHT}, there exists an admissible self-financing strategy $\delta_{\tilde H_T}=\{\delta_{\tilde H_T}(t)=(\delta_{\tilde H_T}^0(t),
	   		\delta_{\tilde H_T}^1(t))^\top,t\in[0,T]\}$ with corresponding stock GOP-denominated price process $\tilde S^{\delta_{\tilde H_T}}_t=\tilde H_t$ given by the  BN pricing formula
	   		\begin{equation}
	   		\tilde H_t={\bf E}^{Q_{S^*}}(\tilde H_{T}|\mathcal{F}_t),
	   		\end{equation} which
	   		delivers the stock GOP-denominated contingent claim	\begin{equation}
	   		\tilde H_T=\tilde S^\delta_T
	   		\end{equation}
	   		$Q_{S^*}$-almost surely at the stopping time $T$.
	   	\end{corollary}
	    Within the set  of admissible self-financing strategies, 	the stock GOP-denominated price 	$\tilde H_t$
	    at time $t\in[0,T]$ coincides with the minimal possible self-financing portfolio process that delivers the stock GOP-denominated contingent claim $\tilde H_T$.\\

	   	\subsection{Hedging Replicable Claims}
	   
	   	 Let us
	   	consider a stock GOP-denominated BN-replicable contingent claim $\tilde H_T$ with a bounded stopping time $T$ as
	   	maturity, where under $Q_{S^*}$ its entire randomness is driven by the  $(Q_{S^*}, \mathcal{\underline F})$-Brownian motion $\bar W$.
	   	   	Each
	   	stock GOP-denominated primary security account  $\tilde S^j_t$,  $j\in\{0,1\}$, is assumed to satisfy an SDE  of the form
	   	\begin{equation}
	   	\frac{d \tilde S^j_t}{\tilde S^j_t}=-\Phi^{j,1}_td\bar W_t
	   	\end{equation}
	   	for $t\in[0,\infty)$ with $\tilde S^j_{0}>0$. We assume that $\Phi^{j,1}_t=\{\Phi^{j,1}_t,t\in[0,\infty)\}$  forms for each $j\in\{0,1\}$ 
	   	a predictable process such that the above stochastic differentials are well defined; see \citeN{KaratzasSh98}. 
	   	For  $t\in[0,\infty)$ we denote by  $\Phi_t=[\Phi^{j,k}_t]_{j,k=0}^{1,1}$ the matrix with elements $\Phi^{j,1}_t$
	   		   	for $j\in\{0,1\}$, and
	   	\begin{equation}
	   	\Phi^{j,0}_t=1
	   	\end{equation}
	   	for $j\in\{0,1\}$.  Let us impose the following assumption:
	   	\begin{assumption} \label{tildeH} We assume that a BN-replicable stock GOP-denominated contingent claim $\tilde H_T$ has for its fair stock GOP-denominated price at time $t\in[0,T]$ under $Q_{S^*}$ a unique martingale representation  of the form
	   		\begin{equation}\label{tildeHT''}
	   		\tilde H_t=\tilde H_{t_0}+\int_{t_0}^{t}x_s d{W_s},
	   		\end{equation}
	   		where $x_t$ is predictable and the integral
	   		\begin{equation}
	   		\int_{t_0}^{t}x_s^2ds<\infty
	   		\end{equation} is $Q_{S^*}$-almost surely finite for 
	   		$t\in[0,\infty)$. 
	   	\end{assumption} 
	   	A stock GOP-denominated self-financing portfolio $\tilde S^\delta_t$ is  a $(Q_{S^*}, \mathcal{\underline F})$-local martingale. In \citeN{Platen25a} the following result is derived:
	   	\begin{theorem}  If the matrix $\Phi_t$ is Lebesgue-almost everywhere invertible, then for a  BN-replicable stock GOP-denominated contingent claim $\tilde H_T$ under Assumption \ref{tildeH} the respective hedging strategy $\delta_{\tilde H_T}$ is given by the vector
	   		\begin{equation}
	   		\delta_{\tilde H_T}(t)=diag({\bf{\tilde S}}_t)^{-1}(\Phi_t^\top)^{-1}\xi_t
	   		\end{equation}
	   		with
	   		\begin{equation}
	   		\xi_t=(-x_t,\tilde H_t)^\top
	   		\end{equation}
	   		for all $t\in[0,T)$.
	   	\end{theorem}
	   	Here $diag({\bf{ S}})$ denotes the diagonal matrix with the elements of a vector ${\bf{ S}}$ as its diagonal. 
	   	     We emphasize that it is the fair price process that delivers at time $T$ the stock GOP-denominated contingent claim $\tilde H_T$ by starting from the most economical minimal possible stock GOP-denominated initial  price $\tilde H_{0}$. Other pricing rules, like the risk-neutral pricing rule, may suggest higher initial prices.\\

	   	\section{BN Pricing under the Minimal Market Model}\setcA
	   	This section illustrates BN pricing for a European put option by employing  the minimal market model (MMM) suggested in \citeN{Platen01a}.
	   	\subsection{Minimal Market Model in Activity Time} 
	   	   	We assume a model where the stock GOP evolves in  some {\em activity time}  $\tau=\{\tau_t, t\in[0,\infty)\}$ with {\em activity}
	   	$
	   	a_t=\frac{d\tau_t}{dt}\in(0,\infty)
	   	$
	   	for $t\in[0,\infty)$ starting with    the  {\em   initial activity time} $\tau_{0}$, where
	   	\begin{equation}\label{tau}
	   	\tau_t=\tau_{0}+\int_{0}^{t}a_sds.
	   	\end{equation} The activity is the derivative of the time in which the stock GOP dynamics evolve. 
	   	The generalized risk-adjusted return $\lambda^*_t$  is a Lagrange multiplier, see Theorem \ref{GOPTheorem}, and does not need to be modeled under BN pricing because it is not relevant under the proposed change of measure. 
	   	For simplicity, we  assume 
	   	that the generalized risk-adjusted return is   proportional to the activity, which means that
	   	\begin{equation}\label{lambda1}
	   	\lambda^*_t=\bar\lambda a_t
	   	\end{equation}  with $\bar\lambda >0$ for $t\in[0,\infty)$.\\

	   	Due to the structure of the SDE \eqref{e.4.1} of the stock GOP, the {\em minimal market model} (MMM) has been suggested in \citeN{Platen01a} as a realistic model for the long-term stock GOP dynamics. The MMM models the stock GOP as a time-transformed squared Bessel process of dimension four, see (8.7.1) in \citeN{PlatenHe06}. We let the MMM evolve in some activity time. 
	   	  For the MMM, described, e.g., in Section 13.2 in \citeN{PlatenHe06},  when evolving in activity time, the  volatility of the stock GOP dynamics in calendar time is given    in the form
	   	\begin{equation}\label{theta2'''}
	   	\theta_{t}=\sqrt{\frac{4e^{ \tau_t}a_t}{S^*_t}}
	   	\end{equation}
	   	for $t\in[0,\infty)$. 
	   	 By applying this model,  the stock GOP is according to \eqref{dS*T} and \eqref{W0} assumed to satisfy under the BN pricing measure the SDE 
	   	\begin{equation} \label{e.4''}
	   	dS^{*}_t
	   	=4e^{\tau_t}d\tau_t+\sqrt{S^*_t4e^{\tau_t}}d\bar W({\tau_t}) \end{equation}
	   	with $S^*_{0}>0$ 
	   	for $t\in[0,\infty)$. 
	   	The process $\bar W({ \tau_{t}})$ represents a  Brownian motion  under $Q_{S^*}$ in activity time with stochastic differential
	   	\begin{equation*}\label{barW0}d\bar W({\tau_t})
	   	=\sqrt{a_t}d \bar W_t 
	   	\end{equation*} 
	   	for $t\in[0,\infty)$.  The SDE \eqref{e.4''} shows that $S^*_t$ represents under $Q_{S^*}$ a time-transformed squared Bessel process of dimension  four; see Definition 1.1 of Chapter XI in \citeN{RevuzYo99} or Equation (8.7.1) in \citeN{PlatenHe06}. Its volatility in activity time exhibits  a leverage effect and is, as a 3/2 volatility model, a constant elasticity of variance model, see, e.g., \citeN{Cox75}, \citeN{Platen97d}, \citeN{Heston97}, and \citeN{Lewis00}.\\ 
	   	\subsection{Observed Activity Time}
	   	By  application of the It\^{o} formula, \eqref{e.4''}, \eqref{barW0}, and \eqref{W0} we obtain the SDE 
	   	\BE \label{e.5}
	   	d\sqrt{S^{*}_t}=\frac{3e^{\tau_t}}{2\sqrt{S^*_t}}d\tau_t+ \sqrt{e^{\tau_t}} d\bar W({\tau_t})
	   	\EE
	   	for $t\in[0,\infty)$. One can observe the  activity time via the quadratic variation of 
	   	the square root of the stock GOP $\sqrt{S^*_t}$. 
	   	Since the measure change from the real-world measure to the benchmark-neutral measure  does not affect the diffusion coefficient in \eqref{e.5}, the quadratic variation of $\sqrt{S^{*}_t}$ becomes
	   	\begin{equation}
	   	[\sqrt{S^{*}_.}]_{0,t}=\int_{\tau_{0}}^{\tau_t}e^{s}ds=e^{\tau_t}-e^{\tau_{0}}
	   	\end{equation}
	   	and the activity time takes the form \begin{equation}\label{tauest}
	   	\tau_t= 	 \ln\left(
	   	[\sqrt{S^{*}_.}]_{0,t}+e^{\tau_{0}}\right)
	   	\end{equation}
	   	for $t\in[0,\infty)$.\\ 
	   	
\begin{figure}[h!]
	   		\centering
	   		\includegraphics[width=14cm, height=6cm]{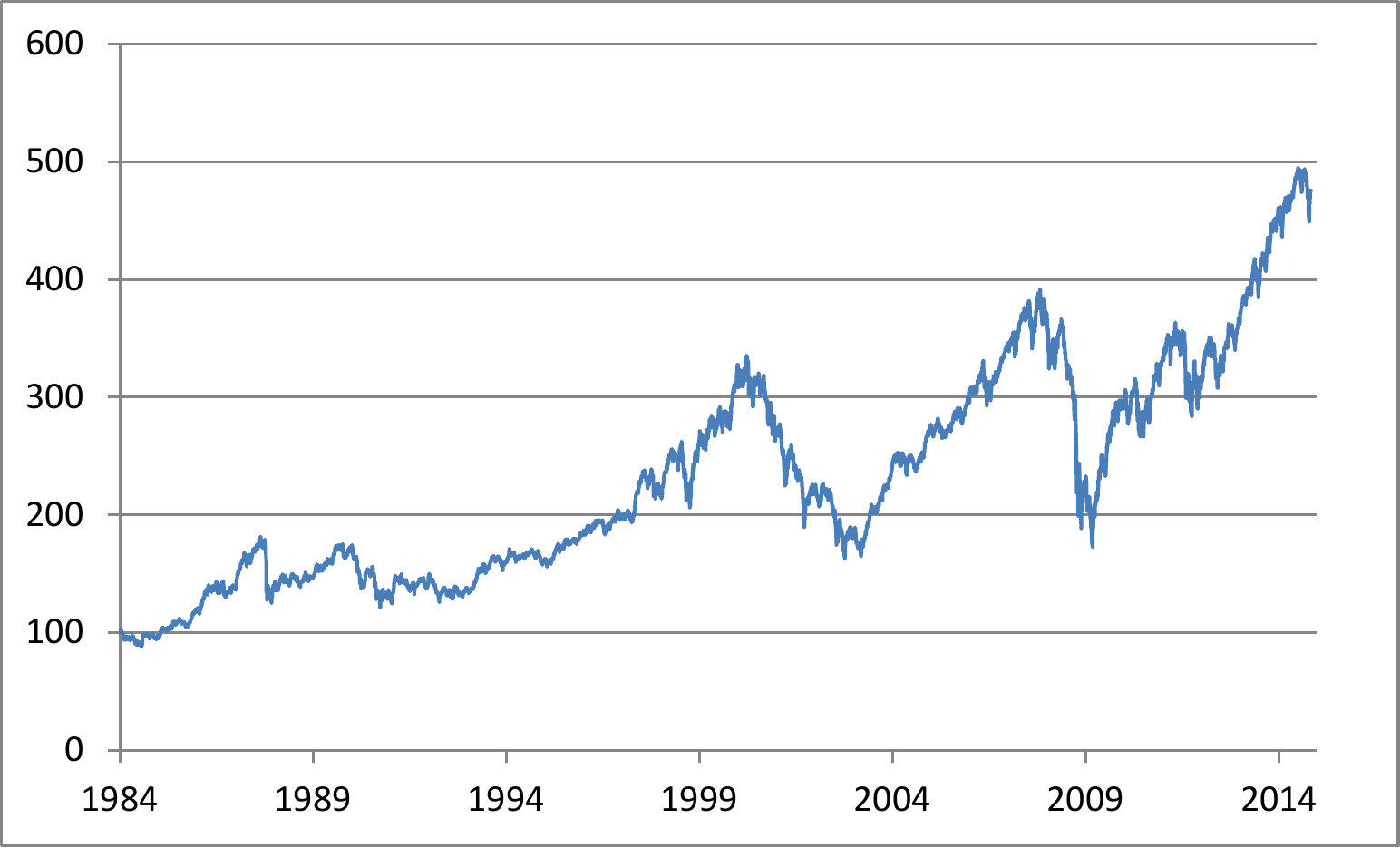}\\
	   		\caption{US Dollar savings account-discounted MCI.}\label{FigAMCI1}
	   	\end{figure}

	   	For  illustration, we consider as underlying security the market capitalization-weighted total return stock index (MCI), where its logarithm is displayed in Figure \ref{FigAMCI1}.  In \citeN{PlatenRe20} the MCI was  generated from stock data  to match closely the daily observed US Dollar savings account-denominated MSCI-Total Return Stock Index for the developed markets. The US Dollar savings account is approximated by a roll-over account of 3-month US T-Bills. We interpret the MCI as a reasonable  proxy for the respective stock GOP. In \citeN{Platen25a} the above-mentioned savings account discounted MCI was employed as a proxy for the stock GOP for pricing and hedging an extreme-maturity zero-coupon bond. We use in the current paper the  parameters fitted in the mentioned study. \\
	   	The approximately linearly evolving activity time is  
	   	approximated by its {\em trendline} \begin{equation}\bar \tau_t=\bar \tau_{0}+ \bar at,
	   	\end{equation}
	which   	is in \citeN{Platen25a} estimated  with
	   	the  {\em slope} $ \bar a\approx 0.053$ and the {\em  initial value} $\bar \tau_{0}\approx 2.15$  for the  period from $t=0 $ at 2 January 1984 until $t=T $ at 1 November 2014. 

	   	\subsection{BN Pricing Measure}

	   	When using  the  trendline $\bar \tau_t$ with constant slope $\bar a\in(0,\infty)$ as a  model for the activity time, it follows by \eqref{sigma10} and  \eqref{theta2'''} that the volatility $\sigma^{S^*}(t)$  of the Radon-Nikodym derivative of the BN pricing measure  equals
	   	\begin{equation}\label{sigma2'}
	   	\sigma^{S^*}(t)=\frac{\bar \lambda \bar a}{\theta_t}=\bar\lambda\sqrt{\frac{\bar aS^*_t}{4e^{\bar{\tau_t}}}}
	   	\end{equation}
	   	for $t\in[t_0,\infty)$.  The following statement is derived in \citeN{Platen25a}:
	   	\begin{theorem}\label{truemartBP}
	   		Under the MMM in activity time with the stock GOP satisfying the SDE \eqref{e.4''}, where the activity time $\tau_t$ is assumed to equal its trendline $\bar \tau_t$, the Radon-Nikodym derivative of the BN pricing measure $Q_{S^*}$ is  a   $(P, \mathcal{\underline F})$-martingale, the measure $Q_{S^*}$ is an equivalent probability measure, and Assumption \ref{RNDMARTINGALE} is satisfied.
	   	\end{theorem}
	   	The martingale property of the Radon-Nikodym derivative $\Lambda_{S^*}(t)$ of the  BN pricing measure $Q_{S^*}$ is secured in the proof of Theorem \ref{truemartBP} by the boundary behavior of the   volatility  $\sigma^{S^*}(t)= \frac{\lambda^*_t}{\theta_t}$, which is by \eqref{theta2'''} proportional to the square root of a squared Bessel process; see, e.g.,  \citeN{MijatovicUr12}, \citeN{HulleyPl12}, and  \citeN{HulleyRu19}.\\

	   	\subsection{BN Pricing of a Put Option}
	   	Under the above model, which employs the trendline of the activity time as a model for the activity time, the transition probability density for the stock GOP $S^*$ under $Q_{S^*}$ has, according to Corollary 1.4 in Chapter XI of \citeN{RevuzYo99}, or equation (8.7.9) in \citeN{PlatenHe06}, the form 
	   	\begin{equation}
	   	p(\bar \tau_t,S^*_t;\bar \tau_s,S^*_s)=\frac{1}{2(e^{\bar \tau_s}-e^{\bar \tau_t})}\left(\frac{S^*_s}{S^*_t}\right)^{\frac{1}{2}}\exp\left\{-\frac{S^*_t+S^*_s}{2(s^{\bar \tau_s}-e^{\bar \tau_t})}\right\}I_1\left(\frac{\sqrt{S^*_tS^*_s}}{e^{\bar \tau_s}-e^{\bar \tau_t}}\right)
	   	\end{equation} for $0\leq t< s<\infty$, where $I_1(.)$ denotes the modified Bessel function of the first kind with index $1$; see, e.g., \citeN{AbramowitzSt72}. Consequently, we know the transition probability density of the stock GOP  under the BN pricing measure, where its key characteristic is the trendline of the  activity time.\\
	   	To illustrate BN pricing and hedging, we consider a European put option  with fixed maturity $T\in(0,\infty)$ and contingent claim \begin{equation}
	   	H_T=\max\left(0,K-S^*_T\right),
	   	\end{equation}
	   which gives the owner at maturity $T$ the right but not the obligation to sell one unit of the MCI for the strike price $K$.	The respective fair  European put option price, initiated at the  time $t=0$, we can write at time $t\in[0,T]$ in the form 
	    \begin{equation}p_{T,K}(t)=S^*_t\frac{ p_{T,K}(t)}{S^*_t}.
	    \end{equation}
	      Since its contingent claim has the form $H_T=\max\left(0,K-S^*_T\right)$, its fair value,  obtained via the BN pricing formula \eqref{BPF}, is given in the denomination of the savings account by the  formula \begin{equation*}
	   	p_{T,K}(t)=S^*_t{\bf{E}}^{Q_{S^*}}\left(\frac{\max\left(0,K-S^*_T\right)}{S^*_T}|\mathcal{F}_t\right)\end{equation*}\begin{equation}\label{put}=K\left(\Psi(x(t);0,\lambda(t,S^*_t))-\exp\left\{-\frac{\lambda(t,S^*_t)}{2}\right\}\right)-S^*_t\Psi(x(t);4,\lambda(t,S^*_t))
	   	\end{equation}
	   	with
	   	   	\begin{equation}
	   	\lambda(t,S^*_t):=\frac{ S^*_{t}}{e^{\bar\tau_T}-e^{\bar\tau_t}},
	   	\end{equation}
	   		   	\begin{equation}
	   	x(t):=\frac{ K}{e^{\bar\tau_T}-e^{\bar\tau_t}},
	   	\end{equation}
	   for $t\in[0,T)$,	and
	   		\begin{equation}
	   		\Psi(x;\delta,\lambda):=P \left(\chi^2_\delta(\lambda)\leq x\right)=\sum_{k=0}^\infty \frac{\exp\{-\frac{\lambda}{2}\}(\frac{\lambda}{2})^k}{k!}\left(1-\frac{\Gamma\left(\frac{x}{2};\frac{\delta+2k}{2}\right)}{\Gamma\left(\frac{\delta+2k}{2}\right)}\right),
	   		\end{equation}
	  where  $\chi^2_\delta(\lambda)$ denotes a non-central chi-square distributed random variable	with $\delta\geq 0$ degrees of freedom and non-centrality parameter $\lambda>0$  for $x\geq 0$; see  equation (13.3.21) in \citeN{PlatenHe06} or Section 3.3.2 in \citeN{BaldeauxPl13}. Here
	   	 \begin{equation}
	   	\Gamma(p)=\int_{0}^{\infty}t^{p-1}e^{-t}dt
	   	\end{equation} 
	   	denotes the gamma function and
	   	\begin{equation}
	   	\Gamma(u;p)=\int_{u}^{\infty}t^{p-1}e^{-t}dt
	   \end{equation} 
	   	the incomplete gamma function for $u\geq 0$ and $p>-1$; see \citeN{AbramowitzSt72} and \citeN{JohnsonKoBa95}.\\
	   	
	   	\begin{figure}[h!]
	   		\centering
	   		\includegraphics[width=14cm, height=6cm]{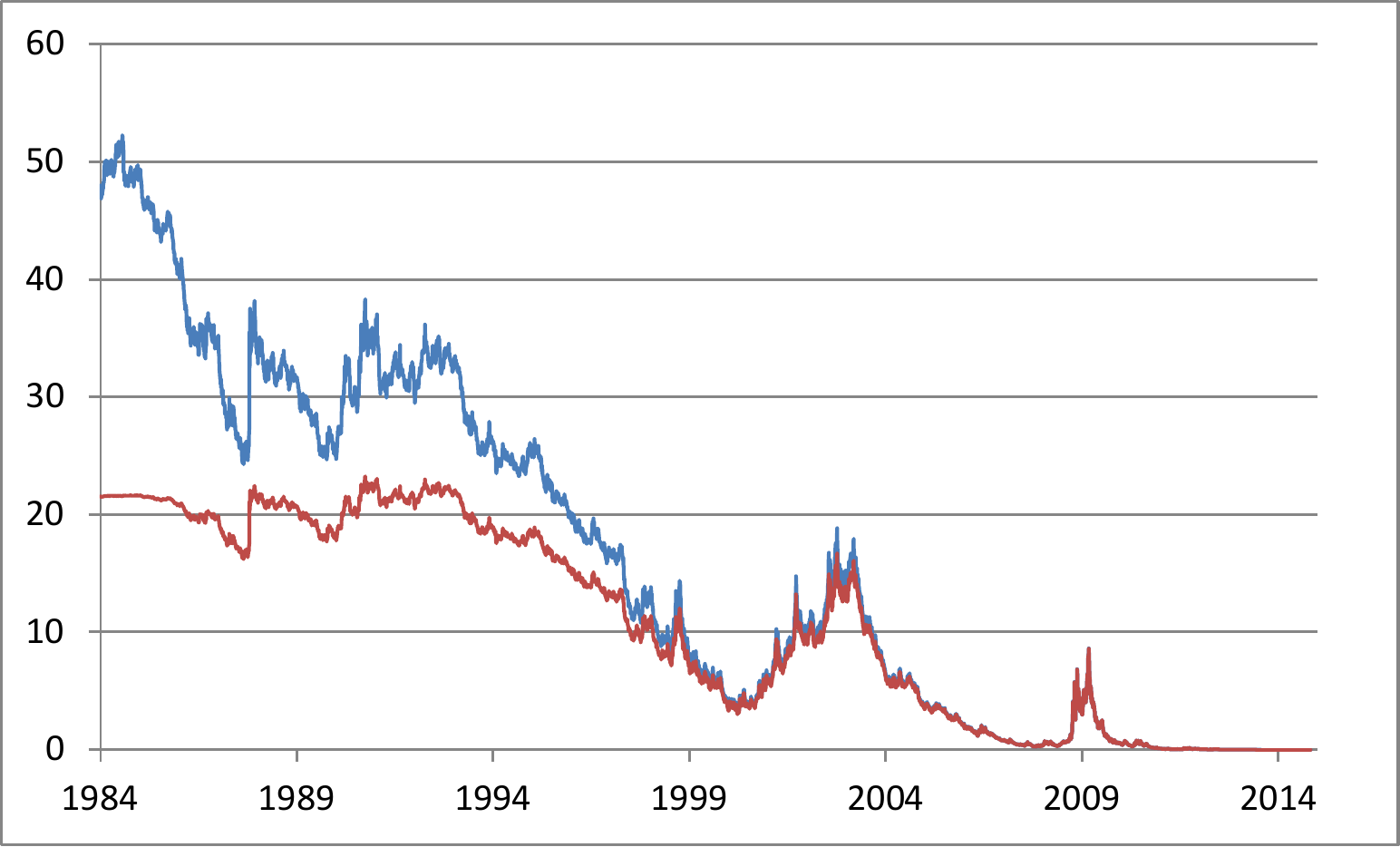}\\
	   		\caption {Fair (red) and risk-neutral (blue)  put option  on the MCI.}\label{FigAMCI3}
	   	\end{figure}
	The option is  initiated at the beginning of our data set with maturity at the end of our observation period. Recall that we approximate the activity time by the respective value of its trendline.    Figure \ref{FigAMCI3} displays in red the savings account-discounted  fair European put option price on the MCI with strike price $K=100$.  One notes that the put option delivers perfectly the targeted payout, where the strike price is significantly lower than the MCI value at maturity. \\
	    
	    The classical formally calculated risk-neutral price for the same dynamics and contingent claim is shown in Figure \ref{FigAMCI3} in blue. It is initially significantly more expensive than the respective fair put price.	It is shown in \citeN{Platen25a} that the putative risk-neutral pricing measure is for the employed MMM not an equivalent probability measure, which makes the respective risk-neutral European put price more expensive than the above fair  price. Similarly to the hedging of a zero-coupon bond illustrated in \citeN{Platen25a}, the fair European put option can be hedged rather accurately over a long period of more than  30 years.  This means, in the given situation there exist  two different self-financing portfolios that hedge the same  put option payoff at the maturity date $T$.  The Law of One Price under the assumptions of classical finance theory does not permit such a situation. The weaker assumption  on the existence of the GOP of the entire market  still permits under the benchmark approach the existence of  these two different portfolios with the same payoff at maturity. 
	   	 Therefore, the original  hypothesis that  risk-neutral put prices on stock indexes are overpriced, which motivated the benchmark approach, cannot be easily rejected when employing the concept of BN pricing and the MMM for pricing an extreme-maturity European put option using historical stock index data. The benchmark approach explains why and how the classical mathematical finance theory could be extended to model more realistically stock markets.

	   	\section*{Conclusion}
	   	The  paper explains and applies the  method of benchmark-neutral  pricing. It employs  the growth optimal portfolio  of  a set of stocks as a num\'eraire and the respective benchmark-neutral pricing measure for pricing.  The  paper assumes the  dynamics of this num\'eraire as those of a squared Bessel process of dimension four in some activity time. For an  extreme-maturity put option on a well-diversified stock portfolio, it is demonstrated that the  benchmark-neutral price is lower  than the respective risk-neutral price. By  extending the  benchmark-neutral pricing methodology, it should be   possible to develop   quantitative methods for    a wide range of contingent  claims relevant to pension and life insurance contracts.

	        \bibliographystyle{\dir chicago}
	        \bibliography{\dir my}
	        
	        \newpage

 \end{document}